# Quantum mechanical current-to-voltage conversion with quantum Hall resistance array

Dong-Hun Chae[1,2], Mun-Seog Kim[1], Wan-Seop Kim[1], Takehiko Oe[3] and Nobu-Hisa Kaneko[3]

[1] Korea Research Institute of Standards and Science, Daejeon 34113, Republic of Korea

[2] The University of Science and Technology, Daejeon 34113, Republic of Korea

[3] National Metrology Institute of Japan, National Institute of Advanced Industrial Science and Technology, Tsukuba, Ibaraki 305-8563, Japan

E-mail: dhchae@kriss.re.kr

## Abstract

Accurate measurement of the electric current requires a stable and calculable resistor for an ideal current-to-voltage conversion. However, the temporal resistance drift of a physical resistor is unavoidable, unlike the quantum Hall resistance directly linked to the Planck constant $h$ and the elementary charge $e$. Lack of an invariant high-resistance leads to a challenge in making small-current measurements below 1 µA with an uncertainty better than one part in $10^6$. In this work, we demonstrate a current-to-voltage conversion in the range from a few nano amps to one microamp with an invariant quantized Hall array resistance. The converted voltage is directly compared with the Josephson voltage reference in the framework of Ohm's law. Markedly distinct from the classical conversion, which relies on an artifact resistance reference, this current-to-voltage conversion does not demand timely resistance calibrations. It improves the precision of current measurement down to $8\times10^{-8}$ at 1 µA.

# 1. Introduction

The precise measurement of small-currents is of prime importance for fundamental research and practical applications. The International System of Units (SI) has been redefined with advances in quantum metrology for the past half-century and the new SI has come into effect since 20 May 2019 [1]. The revision of the SI is based on the philosophy that the SI units are inherently linked to invariant constants of nature via the laws of physics, not to variable physical artifacts. In the revised SI, the ampere is a flow of $1/(1.602\ 176\ 634 \times 10^{-19})$ individual elementary charges per second. Existing single-electron current sources following this definition can hardly increase their current value accurately above 1 nA owing to the absence of a robust correlated quantum mechanical state of matter and the uncontrollable high-order processes [2]. This results in a poor precision of the electrical current in any accessible range far from the uncertainty of a part in $10^9$ by orders of magnitude. In contrast, the voltage and the resistance can be realized stably and reproducibly below a part in $10^9$, stemming from the macroscopic quantum mechanical states of matters, i.e., the weakly coupled superconducting states leading to the Josephson effect [3] and the topologically protected quantum Hall state [4], respectively. For an accurate current, the well-established voltage and resistance quantum standards can be utilized with Ohm's law.

Needs for accurate small-current measurements are ubiquitous. Environmental monitoring of particulate matter and aerosol nanoparticles requires small-current measurements at the femto-ampere level [5]. The current level of radiation dosimetry applications for medical diagnoses and therapies is smaller than 10 pA [6]. Semiconductor industries also require small-current measurements, including leakage current characterization in nanoscale devices from femto-amperes to micro-amperes, an electron-beam intensity of lithography acquired by a Faraday-cup electrometer down to the pico-ampere level, and ionic conduction measurements in organic materials at the pico-ampere level.

In practice, however, achieving electrical current precision smaller than a part in $10^6$ is challenging in most ranges. Conventional methods of current evaluation are categorized as either a capacitor-charging method for current below 1 nA or a voltage-to-resistor method for higher current [7]. In the capacitor-charging method, an accurate current is generated for the current measurement by charging a stable capacitor with a linearly ramped voltage with time. A dominant uncertainty is attributed to a deviation of the capacitance value at the direct current limit from its value determined at the alternative current with a frequency of 1 kHz as well as a drift in time. The uncertainty is consequently larger than a part in $10^5$. For the measurement and generation of a current larger than 1 nA, artifact resistance and voltage references traceable to the quantum Hall and Josephson effects, respectively, are employed, relying on Ohm's law. Uncertainty regarding the current value also arises from the intrinsic temporal drift and the environmental change of artifacts as well as the long calibration chains. Hence, the overall uncertainty typically becomes larger than a part in $10^6$.

Recently, a number of research groups have developed various methods for improving the limitation of the current precision. Brun-Picard and colleagues in the Laboratoire national de métrologie et d'essais (LNE) in France have realized a current source by integrating a quantum Hall resistance, a Josephson voltage reference, and a transformer based on the cryogenic current comparator (CCC) in three separate cryostats [8]. Even though this method requires interconnections with a finite resistance via the ambient environment, it leads to an improvement of precision at the milli-ampere level by two orders of magnitude, compared with the conventional method. Precise evaluation of a high-value resistor is essential for the current-to-voltage conversion in small-current measurements. For instance, a resistance of a thick-film resistor larger than 100 MΩ can change over the measurement uncertainty within a few hours [9]. Giblin and his collaborators at the National Physical Laboratory (NPL) in UK have improved the measurement uncertainty of high-value resistors by frequent calibrations with a CCC resistance bridge [9]. Consequently, the precision of a small-current measurement can be down to a few parts in $10^7$ at the 100 pA level. To reduce the time drift and temperature instability of a high-value resistor, Drung and co-workers at the Physikalisch-Technisch Bundesanstalt (PTB) in Germany have realized a very stable transresistance against time and temperature [10]. PTB's ultra-stable current amplifier (ULCA) comprises a 1000-fold current amplification stage made of a few thousand thin-film resistors to minimize a drift in time and a secondary 1 MΩ current-to-voltage converter. The stability of the overall 1 GΩ transresistance, approximately a part in $10^7$, can be prolonged by a week [10, 11]. This improved approach still cannot fundamentally overcome the intrinsic temporal drift of physical resistors. To achieve an ideal high-value resistance, various arrays have been developed by integrating quantum Hall resistors over the last two decades [see refs. 12, 13 and references therein]. The long-term stability of a 1 MΩ quantized Hall resistance array (QHRA) fabricated at the National Metrology Institute of Japan has recently been investigated systematically [13]. It turned out that the 1 MΩ QHRA is almost invariant for at least approximately 10 months within the measurement uncertainty of a few parts in $10^8$ [13]. Furthermore, the realization of 10 MΩ QHRA and long-term stability are under investigation [14]. This high-value QHRA can bridge the single-electron current source and the Josephson voltage reference through Ohm's law. A consistency test of three electrical quantities realized by different physical effects is known as the quantum metrology triangle experiment [15-17].

Here we demonstrate a current-to-voltage conversion from a few nano amps to one micro-amp with the invariant nature of a 1 MΩ QHRA and the programmable Josephson voltage standard (PJVS). In the absence of an ideal and invariant current source, we employed a stable 1-MΩ-resistance reference comprising stable tens of metal-foil resistors in an ULCA and a stable voltage source. The resistance reference was precisely evaluated before and after the conversion experiment to carry out the applied current estimation between experiments. The 1-MΩ-resistance reference is connected to a 1 MΩ QHRA in series with a stable solid-state voltage source applied in bipolar. By sequentially measuring a voltage drop across the 1-MΩ-resistance reference and the combined Hall voltage in a 1 MΩ QHRA through comparison to PJVS, we have accurately determined the applied current through the circuit by the 1-MΩ-resistance reference and the current via the 1 MΩ QHRA in the framework of Ohm's law. The coincidence between the above two current values within the measurement uncertainty reflects that the QHRA can be exploited for an ideal current-to-voltage conversion for an unknown small-current measurement below 1 μA without frequent resistance calibrations. An Allan deviation study shows that the precision of the current measurement can also be improved beyond that of the classical conversion based on a physical resistor by at least two orders of magnitude.

## 2. Experimental results

### *2.1. Stability of quantum Hall resistance array*

The QHRA was designed by the continued fractional expansion method and realized by integrating 88 quantum Hall devices with the following nominal value, almost 100-fold larger than a single quantum Hall resistance, at filling factor 2 (Fig. 1a inset and 1b)[18]; $R_{\text{nom}} = 10150/131 \times h/2e^2 = 999\ 999.983\ \Omega_{2019}$. Figure 1a shows the quantized Hall resistance plateaus in a QHRA and the corresponding suppression of longitudinal resistance in a single Hall element. A combined Hall resistance close to the nominal value can be accomplished even with experimental challenges, including dissipation and interconnection [18]. Unlike the Josephson array voltage with the zero-resistance superconducting interconnections, QHRA includes classical conductors with a finite resistance comprising interconnecting wires and metal contacts. This non-quantum mechanical contribution of interconnection can be minimized by the triple connection geometry less than a few parts in $10^{10}$[18]. The measured relative deviation from the nominal value is comparable to the measurement uncertainty of 20 parts in $10^9$, as shown in Fig. 1c. More importantly, the quantized Hall resistance is almost invariant within the measurement uncertainty through the thermal cycles for approximately 600 days. This observation is attributed to the experimental realization that the component quantum Hall resistances become dominant in quantity, compared with the classical interconnection resistance in the triple connection geometry. This unchanging nature of the QHRA is markedly distinct from an artifact resistance reference with a temporal drift. Figure 1d depicts a typical time drift of a 1 MΩ physical resistor in an ULCA employed in this work. The temporal drift is smaller than $30 \times 10^{-9}/\text{month}$ as shown in Fig. 1d and the temperature coefficient is smaller than $50 \times 10^{-9}/°C$ [19].

## 2.2. Experimental setup

The experimental setup is illustrated in Fig. 2a. Experiments were performed in a temperature-stabilized laboratory with a temperature change of less than 0.1 ℃ around 23 ℃. To perform a current-to-voltage conversion experiment with the 1 MΩ QHRA, we need an ideal and invariant current source. Owing to the absence of an ideal current source, we have realized a current source by combining the stable 1-MΩ-resistance reference and a stable voltage source. A stable bipolar voltage source consists of a battery-operated solid-state voltage source, a voltage divider, and a relay switch as illustrated in Fig. 2a. We connect the 1-MΩ-resistance reference in an ULCA and the 1 MΩ QHRA in series with the bipolar voltage source connected. We then sequentially measure a voltage drop across the 1-MΩ-resistance reference in the ULCA and the combined Hall voltage of the 1 MΩ QHRA to measure the current in the circuit. Note that an internal potential reference (triangle), provided by an operational amplifier in the ULCA, is used for the voltage measurement of the 1-MΩ-resistance reference in the ULCA [20]. With each voltage probe pair selected by a scanner, a bias voltage is applied in bipolar for a bipolar current driven in the circuit, while the PJVS generates the voltage corresponding to an expected nominal Hall voltage of the QHRA (Fig. 2a). A residual voltage difference is finally acquired by a digital voltmeter for the determination of each voltage. This current-to-voltage conversion accordingly leads to the precision current measurement. The described bipolar measurement is employed to minimize the voltage offset and its temporal drift. Moreover, the polarity of the bias voltage is changed 4 times to compensate for the voltage drift in the measurement circuit (see Supplementary material). During a transient time after the polarity change, the data are not acquired (see the transient behavior in Supplementary material). Therefore, an effective data acquisition time becomes approximately one-third the experimental time. Figure 2b shows a typical measurement example for a nominal applied current of 0.5 µA.

We employed circuit components at the metrology-grade level with a small time drift and a small temperature coefficient. For the bipolar voltage source, a battery-operated 10-V Zener diode (Fluke Model 732B with a time drift smaller than 0.065 part per million (ppm)/day and a temperature coefficient smaller than 0.05 ppm/℃), a voltage divider (Fluke Model 720A with a short-term stability of 0.1 ppm/month and a temperature coefficient of 0.1 ppm/℃), and a homemade switch box with a latching relay (Panasonic Model S2-L2-DC5V) were used. We also utilized a feed-back resistor of a current-to-voltage conversion stage in Magnicon Model ULCA-1 for a stable 1-MΩ-resistance reference. For accurate voltage measurement, a low thermal scanner (Data Proof Model 160B with a thermoelectrical potential smaller than 20 nV) was adopted to minimize the thermal offsets. The voltage difference between the voltage drop across 1-MΩ resistance and the balancing Josephson voltage was measured by an 8½ digit voltmeter (Keysight Model 3458A).

*2.2. Precision measurements of current-to-voltage conversion*

Data sets were acquired for approximately 9 hours to determine the average difference between two current values. We note that the limited holding time of the base temperature near 0.3 K is approximately 9 hours for the single-shot $^3$He refrigerator employed in this work. Figure 3a shows the relative current difference of $(I_{\text{QHRA}} - I_{\text{Ref}})/I_{\text{Ref}}$ with time for a nominal current value of 0.5 µA as an example. Here, $I_{\text{QHRA}}$ and $I_{\text{Ref}}$ are the current values determined by the measured voltage values across the 1 MΩ QHRA and the 1-MΩ-resistance reference in the ULCA, respectively. The inset shows a histogram of the relative difference. The histogram can be fitted reasonably well by the Gaussian function. This might imply that uncorrelated random noise is dominant in the measurement.

To investigate the noise characteristics and the ultimate measurement resolution of the measurement setup, the Allan deviation [21] of the relative difference is plotted with the time trace data set. Three representative Allan deviation plots are depicted in Fig. 3b. Traces for relatively large nominal current values of 0.5 μA and 100 nA are described well by the inverse square root time dependence. This indicates that the uncorrelated white noise is predominant in the measurement. The statistical uncertainty reaches a few parts in $10^8$ for 0.5 μA, as shown in Fig. 3b. This reflects that the uncertainty would become even smaller if a longer measurement is allowed in the measurement setup. The uncertainty determined in this work is not the fundamental limit but the experimental limit due to the finite holding time of the base temperature. For smaller current such as 10 nA, the white noise characteristics of the Allan deviation are prolonged by half an hour of sampling time. After that, the typical flicker noise characteristics with zero slope emerge. Additionally, an injection current [22] from a digital voltmeter may play a noticeable role at the smaller current (see Supplementary material), which is not the white noise, to be discussed later. This means that the measurement duration is not the limiting factor for precision in the smaller current regime.

Figure 4 summarizes our key results on the relative current difference and the measurement uncertainty in the current range from 5 nA to 1 μA. For comparison, the calibration and measurement capability (CMC) values of PTB registered on KCDB in the International Bureau of Weights and Measures (BIPM) is plotted with black circles in the investigated current range [23]. Square, pentagon, and diamond symbols depict the measurement uncertainties for the ULCA (PTB)[11], the measurement of single electron pump with a timely calibrated resistor (NPL)[9], and the programmable quantum mechanical current generator (LNE)[8], respectively. The x-axis is the nominal current value driven in the circuit. The red hexagon and green square symbols represent the relative difference and uncertainty, respectively. The relative difference is smaller than the uncertainty. This indicates that the current values measured by the 1-MΩ-resistance reference and 1 MΩ QHRA coincide with each other within the uncertainty. This reflects the fact that the 1 MΩ QHRA can be utilized for an ideal current-to-voltage conversion. The achieved uncertainty at the nominal current of 1 μA is smaller than one part in $10^7$. This value is smaller than that of the CMC based on an artifact resistance reference by at least one order of magnitude. The uncertainty increases logarithmically with decreasing current. Detailed uncertainty analysis is presented in the next section.

## 2.3. Uncertainty budget

The relative current difference ($r_I$) is expressed as follows;

$$r_I \equiv \frac{I_{\text{QHRA}} - I_{\text{Ref}}}{I_{\text{Ref}}} = \frac{\frac{V_{\text{QHRA}}}{R_{\text{QHRA}}} - \frac{V_{\text{Ref}}}{R_{\text{Ref}}}}{\frac{V_{\text{Ref}}}{R_{\text{Ref}}}} = \frac{\frac{R_{\text{Ref}}}{R_{\text{QHRA}}}}{\frac{V_{\text{Ref}}}{V_{\text{QHRA}}}} - 1$$

The uncertainty contribution for each physical parameter ($x_i$) is obtained by multiplying the sensitivity coefficient ($c_i = \frac{\partial r_I}{\partial x_i}$) and the standard error ($u_i$) of $x_i$, taking into consideration the probability distribution factor of $x_i$. The probability distribution factors of the rectangular probability distribution for the systematic type-B uncertainty and the normal probability distribution for the statistical type-A uncertainty are $1/\sqrt{3}$ and 1, respectively. The measurement uncertainty is calculated as follows;

$$u_c = \sqrt{\sum_i c_i^2 u_i^2}$$

Table 1 summarizes important uncertainty contributions for nominal current values of 1 μA and 10 nA. The uncertainty for the 1 MΩ resistance calibrated by a CCC resistance bridge is described in detail in a previous report [13]. The corresponding relative uncertainty is smaller than 20×10$^{-9}$. The limited resolution (10 nV) of the digital voltmeter (DVM) (Keysight 3458A) for the voltage measurement may also introduce uncertainty. The corresponding relative uncertainties for 1 μA and 10 nA become 10×10$^{-9}$ and 1×10$^{-6}$, respectively. The finite input impedance of the used DVM introduces an error in the voltage measurement. We evaluated the input resistance [24]. It is approximately 10 TΩ. For the direct voltage measurement by the DVM, the relative uncertainty becomes 0.1×10$^{-6}$ for the measurement of the voltage drop across the 1-MΩ resistor. However, since the input voltage is typically reduced down to tens of μV by the balancing PJVS, the relative uncertainty is reduced accordingly by orders of magnitude. For instance, the relative uncertainties of the input impedance for 1 μA and 10 nA become 1×10$^{-11}$ and 1×10$^{-9}$, respectively. Finally, the statistical type-A uncertainties of the voltage ratios become 76×10$^{-9}$ and 4.0×10$^{-6}$, respectively. The measurement uncertainty is calculated by taking into account the dominant uncertainty contributions as shown in Table 1.

## 3. Discussions

The precision of the current-to-voltage conversion in this experiment may not be the fundamental limit. Two improvements for measurement can be conceived. For measurement of a voltage from high impedance, a charge injection [22,23] from the input circuitry of a digital voltmeter needs to be taken into account. The temporal behaviour of the injection current of the employed digital voltmeter is plotted in Supplementary material. To suppress an excess voltage in high resistance induced by the injection current at the pico-ampere level, a differential nanovoltmeter with an input chopper [25] can be employed for future experiments. Another experimental factor for the limited precision is the data acquisition time stemming from the restricted operational time of the single-shot $^3$He cryostat used. Therefore, the statistical uncertainty can be reduced according to the Allan deviation plot in Fig.3b if a longer measurement is allowed with a $^3$He cryostat that is operational in a continuous flow mode.

Quantized Hall resistance in array does not require timely calibrations for the current-to-voltage conversion according to the stability investigation even though the measurement uncertainty is comparable with that of timely calibrated ULCA [11], which is a state-of-the-art current amplifier based on artefact resistor networks with an intrinsic time drift. Moreover, QHRA can be integrated in the cryogenic environment with other quantum mechanical elements, such as the Josephson voltage and the cryogenic current comparator amplifier, for fundamental metrological research. For instance, the integration of QHRA and PJVS for a quantum mechanically enhanced current source can be envisaged. That is, a voltage-to-current conversion through the QHRA can be realized. In Fig. 2a, if the 1-MΩ-resistance reference and 1 MΩ QHRA are interchanged and the bipolar voltage source is replaced by another PJVS, an experiment on a quantum current source can be performed. With the QHRA and PJVS possibly integrated in one cryostat for a lower thermal noise, a quantum mechanical current source can be realized without the CCC amplifier unlike the previous work by the LNE [8]. A small-current below 1 µA can be generated with 1 MΩ and a higher value QHRA. Also, the parallelization of multiple quantum Hall resistors would allow a larger current possibly up to the milliampere level.

## 4. Conclusions

In conclusion, we have experimentally demonstrated a current-to-voltage conversion with the invariant 1 MΩ quantized Hall array resistance and the programmable Josephson voltage standard. The lack of an ideal current source makes us employ a stable high-value resistance reference and a solid-state voltage source at ambient temperature both traceable to the quantum Hall and Josephson effects, respectively to realize a stable current source for the conversion experiment. By comparing an applied current through the 1-MΩ-resistance reference in ULCA and a current value via the 1 MΩ quantized Hall array resistance, calculated by taking the ratio of the measured Hall voltage to the array resistance, we have determined a ratio for the current-to-voltage conversion. Here, we measured the converted Hall voltage through direct comparison with the programmable Josephson voltage standard. Two current values coincide with each other within the measurement uncertainty in the investigated range from 5 nA to 1 µA. This indicates that a high-value quantized Hall array resistance can be utilized for a current-to-voltage conversion, which is quantum mechanically enhanced by the quantum Hall effect, without timely resistance calibrations. It is conspicuously distinguishable from the conventional current-to-voltage conversion. Additionally, the precision of current measurement based on this current-to-voltage conversion is achieved down to $8\times10^{-8}$ at 1 µA. Moreover, the demonstrated precision, as well as the value of realized quantum Hall resistance array, is not the fundamental limit.

**Acknowledgements**

The authors thank Dietmar Drung, Hansjörg Scherer, Martin Götz, and DongSu Lee for valuable discussions. The authors are also grateful to YongKi Park for making a photograph of the array. This research was supported by the "Enhancement of Measurement and Standards Technologies in Physical SI units" (KRISS-2019-GP2019-0001) and the "Research on Redefinition of SI Base Units" (KRISS-2019-GP2019-0003) funded by the Korea Research Institute of Standards and Science. This work was partially supported by "Graphene Impedance Quantum Standard" (NRF-2019K1A3A1A78077479) funded by National Research Foundation of Korea. This work was also supported by JSPS KAKENHI Grant Numbers JP18H05258 and JP18H01885. The QHRA device was fabricated at the AIST Nano-Processing Facility, supported by "Nanotechnology Platform Program" of the Ministry of Education, Culture, Sports, Science and Technology (MEXT), Japan.

D.-H.C., W.-S.K., T.O. and N.-H.K. conceived the experiments. D.-H.C. designed the experimental approaches. T.O. fabricated and pre-characterized the quantum Hall resistance array. M.-S.K. D.-H.C. and W.-S.K. performed the metrological measurements. D.-H.C. analyzed the data and the uncertainty. All authors discussed the results. D.-H.C. wrote the manuscript with extensive comments made by all the other authors.

**Figure 1**

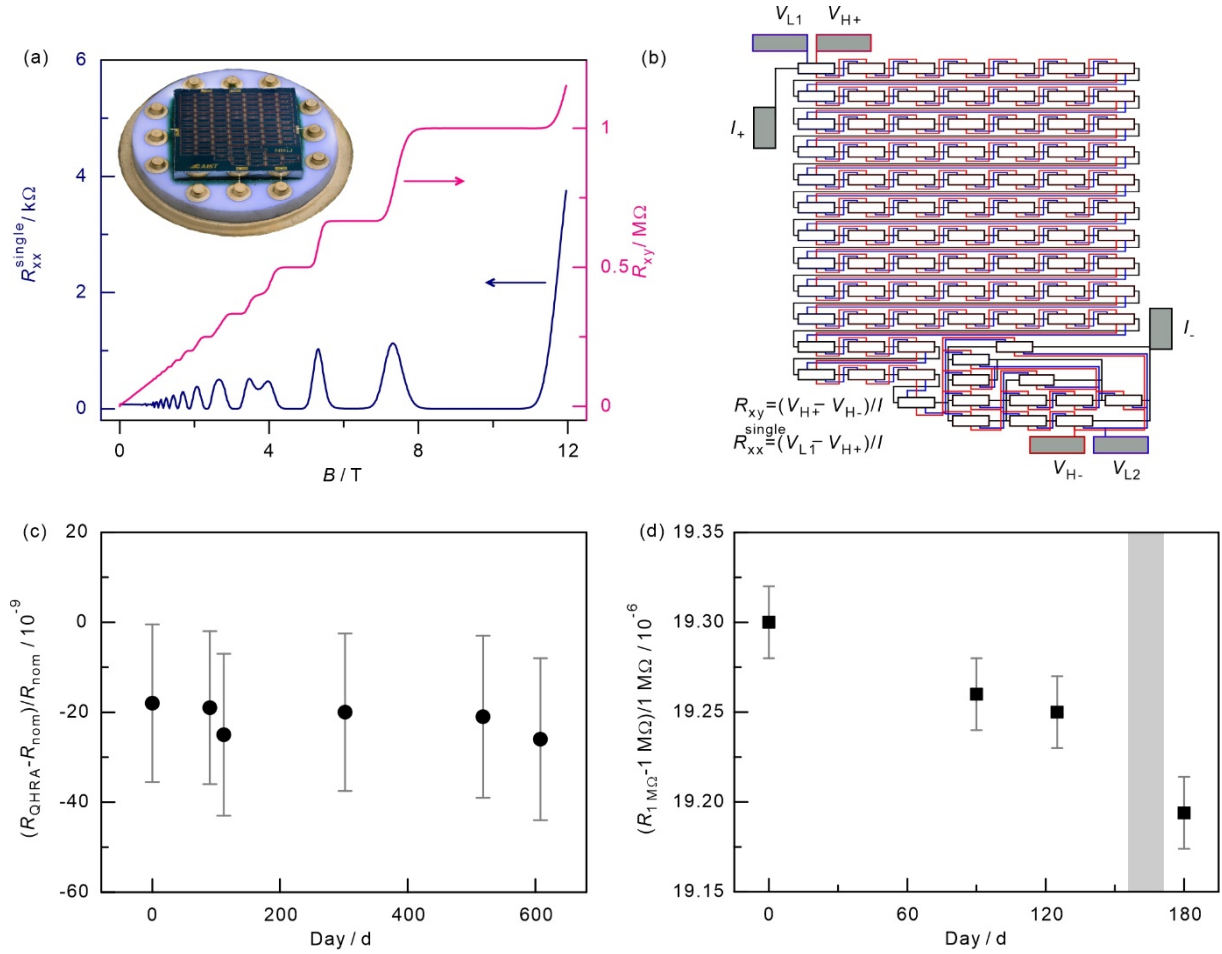

**Figure 1.** Magnetotransport and long-term stability of quantum Hall resistance array. (a) Combined Hall resistance of the array and longitudinal resistance of a single Hall element as a function of the magnetic field at 0.3 K with a 1 μA current. Inset shows a photograph of the measured 1 MΩ quantum Hall resistance array. (b) Schematic circuit diagram of the array. (c) Long-term stability of the quantized Hall resistance in the thermally cycled array. The y-axis is the relative deviation of the quantized Hall resistance at 9.5 T corresponding to the filling factor 2 from the designed nominal value. The first three data points are taken from a previous work [13]. (d) Temporal drift of a stable 1-MΩ-resistance reference in an ULCA at 23 ℃ employed in the experiment. The y-axis represents the relative resistance difference from 1 MΩ. The shaded region indicates the experimental period of the subsequent measurements.

**Figure 2**

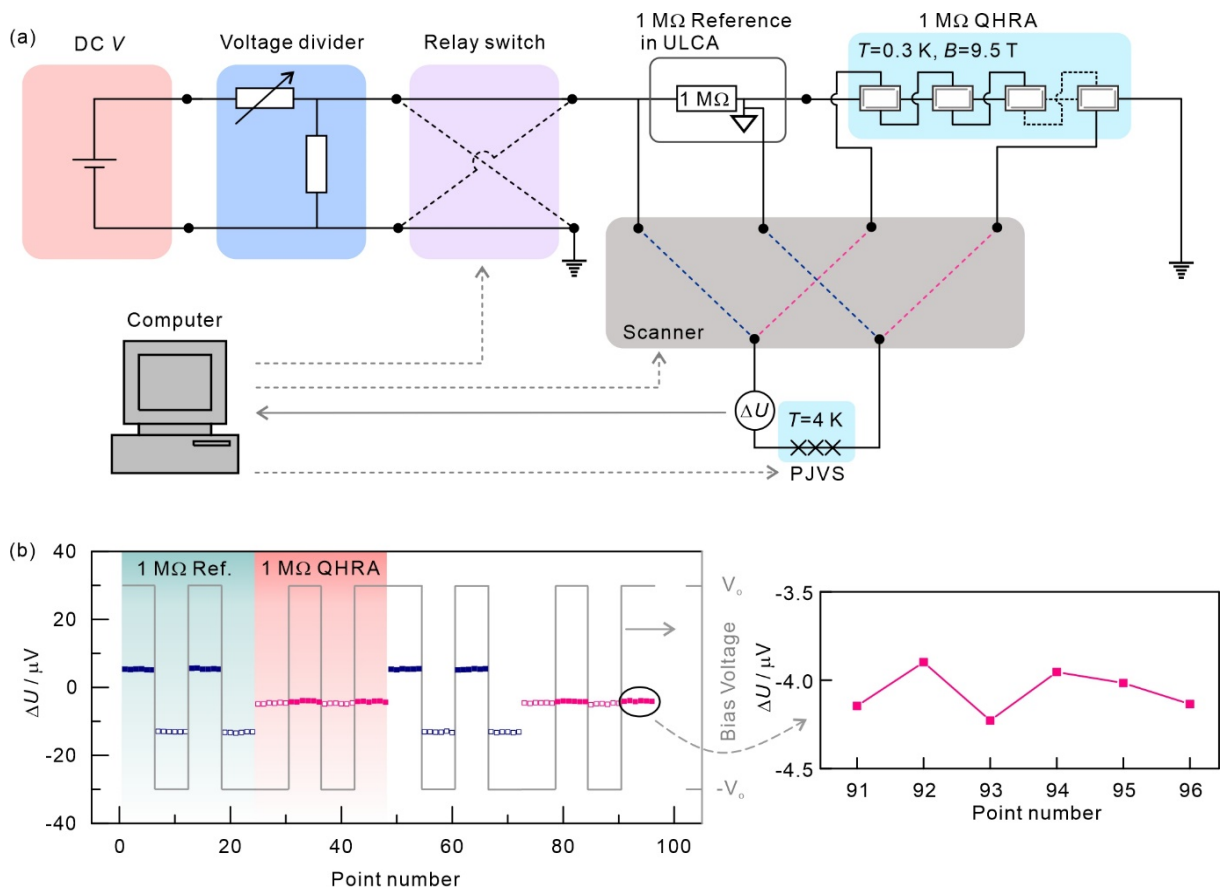

**Figure 2.** Experimental setup and bipolar voltage measurement. (a) A 1-MΩ-resistance reference in ULCA, 1 MΩ quantum Hall resistance array, and bipolar voltage source are connected in series. The bipolar voltage source consists of a battery-operated solid-state voltage source, a voltage divider, and a relay switch. For voltage measurement, a programmable Josephson voltage reference, a digital voltmeter, and a scanner are configured as illustrated in the measurement circuit. (b) For accurate voltage measurement, the voltage difference (Δ$U$) between a voltage drop across 1-MΩ-resistance reference (/1 MΩ quantized Hall array resistance) and a balancing Josephson voltage corresponding to an expected nominal Hall voltage of the quantum Hall resistance array is acquired by a digital voltmeter, with the bias voltage applied in bipolar. The re-scaled data trace is depicted on the right side.

**Figure 3**

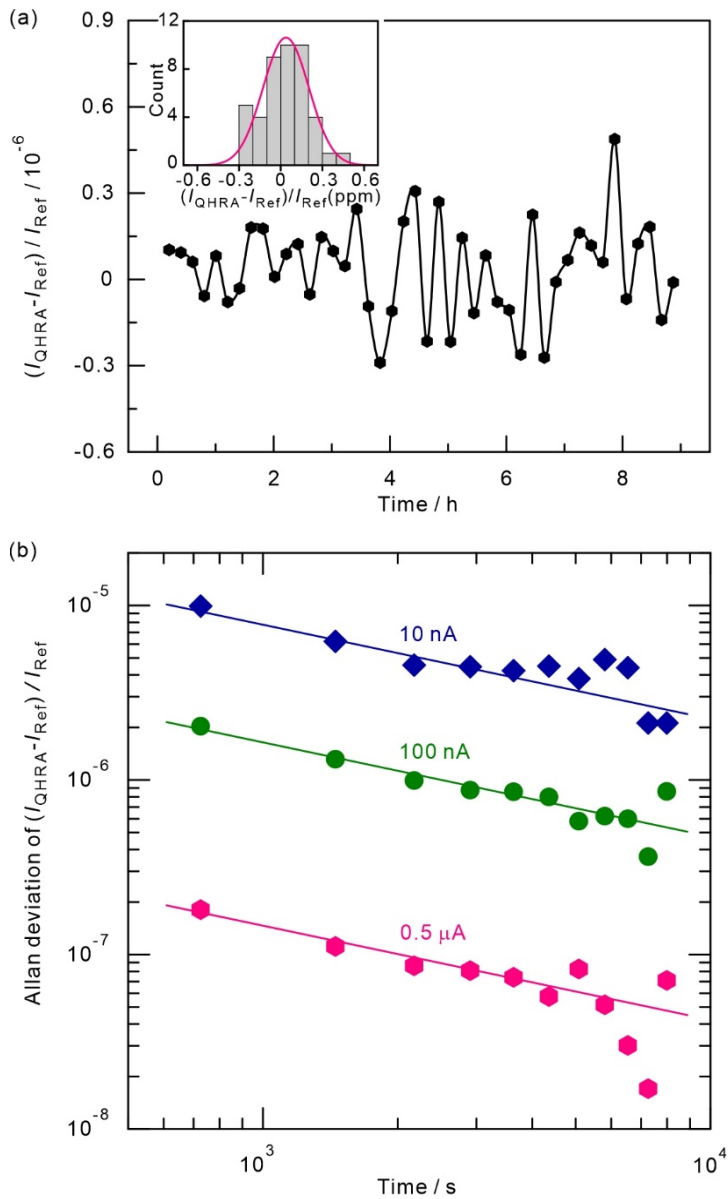

**Figure 3.** Time trace and Allan deviation of relative current difference. (a) Time trace of relative current difference between current values determined by sequentially measured voltage values for 0.5 μA nominal current. Inset exhibits a histogram of the time trace fitted by the Gaussian function. (b) Allan deviations of the relative current difference for nominal values of 0.5 μA, 100 nA, and 10 nA. The fitted inverse square root time dependence ($1/\sqrt{\tau}$) overlaps each Allan deviation.

**Figure 4.**

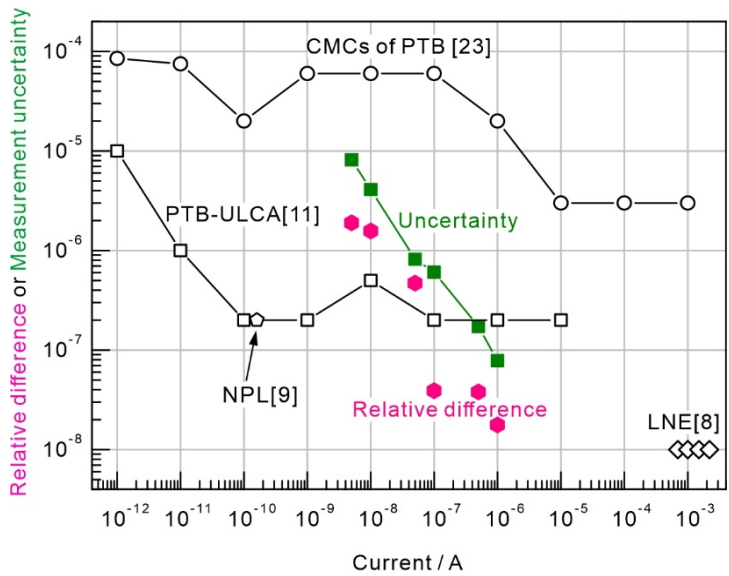

**Figure 4.** Relative current difference and measurement uncertainty. Red hexagons show the relative current difference of $(I_{\text{QHRA}} - I_{\text{Ref}})/I_{\text{Ref}}$. Green squares exhibit the measurement uncertainties. The CMC values of PTB registered in the BIPM are plotted by black circles for comparison. Square, pentagon, and diamond symbols depict the measurement uncertainties for the ULCA (PTB)[11], the measurement of single electron pump with a timely calibrated resistor (NPL)[9], and the programmable quantum mechanical current generator (LNE)[8], respectively.

**Table 1.** Uncertainty budget for relative current difference.

|  | Nominal current values | |
|---|---|---|
|  | $I= 1\ \mu A$ | $I= 10$ nA |
| Contribution | Relative Uncertainty | Relative Uncertainty |
| Resistance measurement | $20\times10^{-9}$ | $20\times10^{-9}$ |
| Resolution of DVM | $10\times10^{-9}$ | $1\times10^{-6}$ |
| Input impedance of DVM | $1\times10^{-11}$ | $1\times10^{-9}$ |
| Type-A | $76\times10^{-9}$ | $4\times10^{-6}$ |
| Measurement uncertainty ($k$=1) | $7.8\times10^{-8}$ | $4.1\times10^{-6}$ |